\documentclass[twocolumn, pre, floatfix]{revtex4-2}
\usepackage{bm}
\usepackage{color}
\usepackage{graphicx}
\graphicspath{{Figures/}}

\usepackage{amsfonts}
\usepackage{amsmath}
\usepackage{mathtools}
\usepackage{hyperref}
\usepackage{verbatim}
\usepackage{xfrac}
\usepackage{epstopdf}

\def\be{\begin{equation}}
\def\ee{\end{equation}}

\begin{document}



\title{Emergence of Hexanematic Order in a Growing Confluent Cell Monolayer}

\author{Hon Lin Too\textit{$^{a}$}}
\author{Farisan Dary\textit{$^{a}$}}
\author{Isaac Si Yuan Ling\textit{$^{a}$}}
\author{Dustin Erhard Theofilus\textit{$^{a}$}}
\author{Duo Zhang\textit{$^{b}$}}
\author{Haiyi Liang\textit{$^{b}$}}
\author{Ee Hou Yong\textit{$^{a,}$}}
\email{eehou@ntu.edu.sg}
\affiliation{\textit{$^{a}$}Division of Physics and Applied Physics, School of Physical and Mathematical Sciences, Nanyang Technological University, 637371,Singapore}
\affiliation{\textit{$^{b}$}CAS Key Laboratory of Mechanical Behavior and Design of Materials, Department of Modern Mechanics, University of Science and Technology of China, Hefei, Anhui, 230026, China}





\begin{abstract}
Collective migration of epithelial layers underlies processes ranging from wound healing to cancer invasion. A defining yet challenging feature is the emergence of distinct cell morphologies within a single migrating confluent sheet, with larger, elongated cells at the active boundary and smaller, hexatically ordered cells in the bulk. Here, we develop a stochastic particle–Voronoi framework that captures this hexanematic organization without prescribing target geometries and distinct cell types. We show that boundary-driven collective motion generates an outward velocity gradient. This gradient, coupled to a density and velocity-dependent division rule, produces peripheral cells that are larger, more elongated, and more defect-prone, while bulk cells remain smaller, isotropic, and hexagonally packed. These results show how minimal, non-equilibrium mechanical interactions give rise to emergent, self-organized tissue-scale patterning during collective migration.
\end{abstract}

\maketitle
\section{Introduction}


Collective cell migration is a fundamental biological process central to tissue morphogenesis, wound healing, and cancer progression. Unlike single-cell motility, collective migration emerges from the coordinated movement of cells that remain mechanically coupled to their neighbors. This coupling enables tissues to migrate as cohesive units while still accommodating local variability, giving rise to emergent properties such as long-range force transmission \cite{tambe2011collective} and large-scale pattern formation \cite{lander2011pattern, campas2009shape, manukyan2017living}. Because of its prevalence across biological contexts, collective migration has been the focus of extensive experimental and theoretical investigations aimed at uncovering how single-cell properties integrate into tissue-scale dynamics.

\begin{figure*}[htb!]
    \centering
	\includegraphics[width=\textwidth]{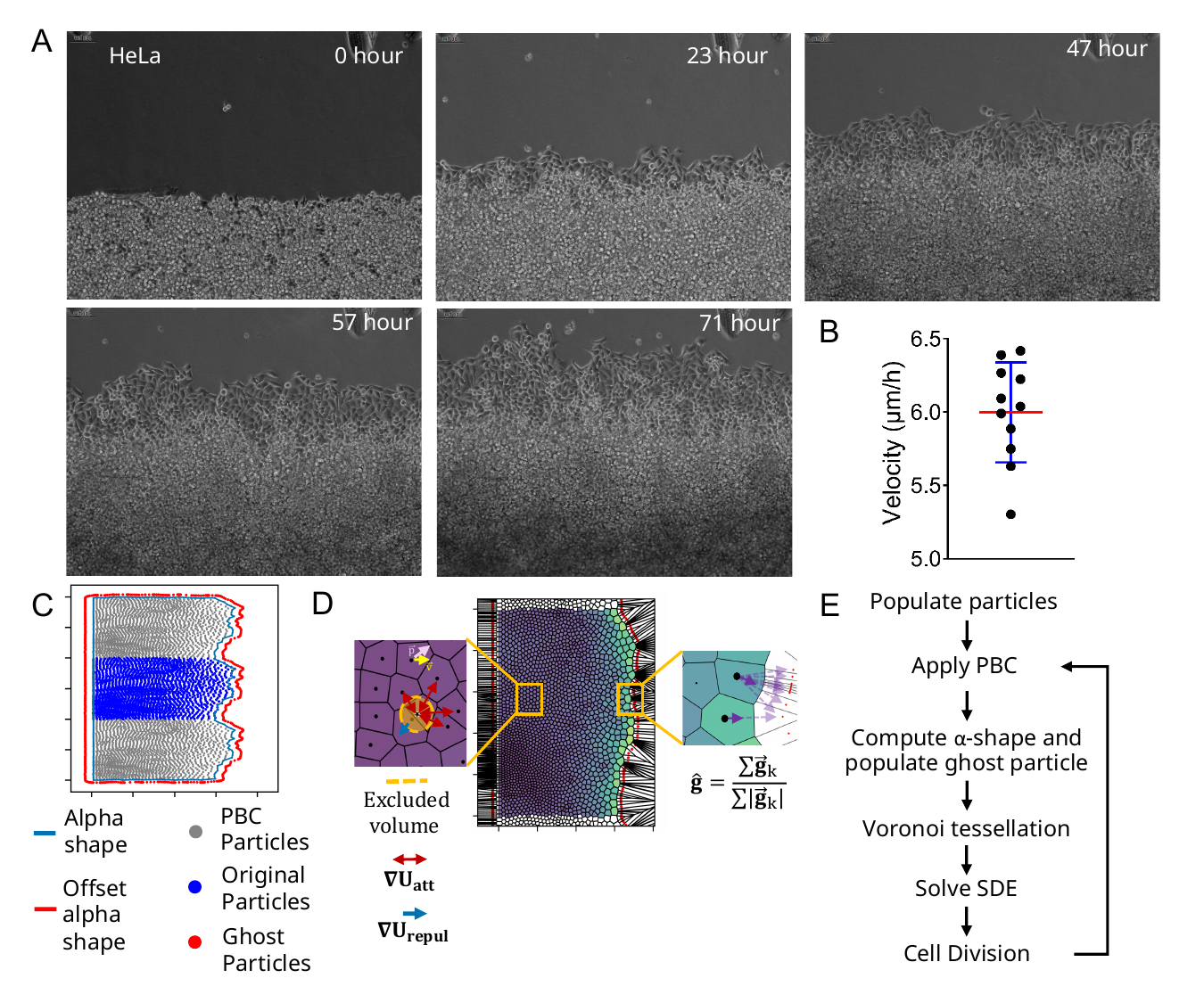}
	\caption {Overview of our computational model.
({\it A}) Time-lapse images of wound closure in a HeLa cell monolayer: peripheral cells are larger and elongated, while bulk cells are smaller and more regular. 
({\it B}) Boxplot of the border speed measured across 11 independent experiments, with a mean value of 6.0$\mu$m/hr.
({\it C}) Particle distribution before Voronoi tessellation. Blue: real particles; grey: periodic copies. The alpha shape defines the boundary, which is offset outward to place ghost particles.
({\it D}) Illustration of particle dynamics. Cell velocity and motility vector are updated through interactions with Voronoi neighbors. 
({\it E}) Simulation workflow.
}
\label{fig:1}
\end{figure*}

A key feature of collectively migrating sheets is the spatial segregation of cell size and morphology. Peripheral cells at the leading edge are typically larger and more elongated (i.e., polarized), whereas bulk cells in the interior remain smaller, more compact, and more isotropic and hexatic \cite{yang2016probing, richardson2016leader, mayor2016front, trepat2018mesoscale, matsuzawa2018alpha, pasupalak2020hexatic}. We refer to this coexistence of the hexatic-dominant and nematic-dominant regions as hexanematic separation. Hexanematic separation has been observed in many systems, including HeLa cell monolayers (Fig.~\ref{fig:1}{\it A}), in zebrafish gastrulation \cite{li2025emergence} and the formation of oncostreams in glioma \cite{comba2022spatiotemporal}. These differences are not merely morphological but are clearly correlated with cell dynamics. Both theoretical and experimental evidence show that more polarized, nematically ordered cells often migrate faster and exhibit a fluid-like (unjammed) phase, whereas hexatically ordered cells tend to move more slowly or remain largely immobile, corresponding to a jammed phase \cite{park2015unjamming, ladoux2016front, bi2016motility}. 

Importantly, cell shape is inherently dynamic, continuously remodeled by mechanical forces transmitted through intercellular junctions that regulate how cells pack, deform, and divide \cite{devany2021cell, eckert2023hexanematic, manukyan2017living, notbohm2016cellular, gov2011moving}. Numerous experiments have documented heterogeneous stress distributions within epithelial monolayers \cite{trepat2009physical,  trepat2011plithotaxis, tambe2011collective, brugues2014forces, malinverno2017endocytic}. Such mechanical feedback drives morphological changes \cite{yang2017correlating, armengol2024hydrodynamics}, notably during wound healing \cite{fenteany2000signaling, brugues2014forces}.  
Consistent with this view, a recent study reported that E-cadherin knockout MDCK-II monolayers, characterized by reduced cell-cell adhesion, exhibit lower hexatic order and packing density than wild type\cite{eckert2023hexanematic}. 
This transition is likely driven by weakened adhesion, increased cell size, and enhanced cellular motility. These observations suggest that the emergence of hexanematic separation is fundamentally a problem of active many-body mechanics. Addressing this question requires theoretical models capable of coupling collective motion, cell growth, and evolving cell morphology.

Existing models of confluent tissues can generally be divided into three categories---particle-based models (PBMs), vertex models (VMs), and self-propelled Voronoi (SPV) models---each offering different strengths and trade-offs \cite{vicsek2012collective, alert2020physical, basan2013alignment, sepulveda2013collective, chepizhko2018jamming, bi2015density, fletcher2014vertex, bi2016motility, li2014coherent, barton2017active, yang2017correlating, armengol2024hydrodynamics}. PBMs efficiently capture collective migration through interacting active particles but lack explicit cell geometry \cite{basan2013alignment, sepulveda2013collective, chepizhko2018jamming}. VMs explicitly resolve cell morphology through polygonal tessellations but require prescribed target geometries \cite{bi2015density, fletcher2014vertex, barton2017active}.SPV models couple active particle motion to Voronoi tessellations, naturally linking cell shape and motility, although morphology remains governed by an underlying shape energy \cite{bi2016motility, yang2017correlating, armengol2024hydrodynamics}. Existing models successfully reproduce either collective migration or cell-shape transitions, yet few provide a minimal framework capable of simultaneously capturing the spontaneous spatial segregation of cell morphology, orientational order, and motility observed in experiments.

We hypothesize that hexanematic separation arises from differential velocity within the tissue: where local crowding suppresses mobility, cells remain densely packed and hexatically ordered; conversely, in regions of higher mobility, reduced crowding permits growth and cells unjam into a more fluid, nematically ordered state. Critically, this heterogeneity is not imposed, cells throughout the tissue share identical mechanical and dynamical rules, and the spatial segregation into hexatic and nematic regions emerges spontaneously from growth and mobility, seeded by an active boundary at the tissue edge. This stands in contrast to existing vertex and self-propelled Voronoi models, in which distinct cell populations or target-shape parameters must typically be prescribed to generate comparable heterogeneity. To formalize this hypothesis, we introduce a two-dimensional particle–Voronoi framework that builds on the stochastic interacting-particle model of Sepúlveda et al. \cite{sepulveda2013collective}: cells are represented as active particles whose positions define a Voronoi tessellation, so that cell shape, area, and neighbor topology emerge directly from particle configurations rather than from a prescribed shape energy. This coupling allows us to track how local mechanical interactions and division events sculpt cell geometry in real time, providing a minimal setting in which to test whether mechanical interactions and kinetic asymmetry due to the active boundary are sufficient to reproduce the hexanematic organization observed in HeLa monolayers. The goal of this model is not to quantitatively fit a specific experimental system, but to establish that a minimal set of local mechanical rules is sufficient to generate hexanematic separation as an emergent, robust phenomenon. Therefore, any comparison with experiments are primarily at the level of qualitative trends and characteristic magnitudes, rather than direct parameter fitting.

\section{Stochastic particle-Voronoi model}

To investigate hexanematic separation in a confluent cell sheet, we modify the stochastic differential equations (SDEs) of Sep{\'u}lveda model \cite{sepulveda2013collective} by coupling the velocity of the $i$-th cell, $\mathbf{v}_i$, with a time-dependent motility vector $\mathbf{m}_i$. The motility vector represents the active self-propulsion term, introducing persistent, internally generated motion arising from chemical and mechanical signaling within and between cells. Our model is given by
\begin{align}
\frac{d\mathbf{v}_i}{dt} &= -\alpha_1 \mathbf{v}_i + \kappa \mathbf{m}_i + \sum_{j\in n_i }\nabla U(\mathbf{r}_{ij}, A_i),\\
\frac{d\mathbf{m}_i}{dt} &= -\alpha_2 \mathbf{m}_i + \sigma_i \pmb\eta_i + \beta\sum_{j\in n_i}(\mathbf{m}_j - \mathbf{m}_i) + D\mathbf{g}_i.
\end{align}
where $\alpha_1$ and $\alpha_2$ are damping coefficients, $\beta$ controls motility alignment between neighbors, $\kappa$ couples velocity to intrinsic motility and $n_i$ is the set of Voronoi neighbors of the $i$-th cell. 

In addition, we augment this model with explicit cell shape information via Voronoi tessellation, combining the flexibility of particle-based formulations with a geometry-resolved description of cells. At each simulation step $t$, cell shapes (and areas $A_i(t)$) are obtained from the Voronoi tessellation, while the tissue boundary is defined by the alpha shape of the particle ensemble (Fig.~\ref{fig:1}{\it C}). Our area-dependent formulation allows cells to dynamically adapt their areas in contrast to vertex and SPV models that impose fixed target areas and perimeters. This adaptive growth mechanism is crucial for generating differential cell size in the model. Cell division is implemented through a minimum division radius, $R_{d}$, which specifies the minimum size required for division and serves as a key control parameter for the division rate. Division therefore functions as the primary size-regulation mechanism, preventing unchecked growth.

The interaction potential is a multivariable function of both neighbour separation, $\mathbf{r}_{ij}(t)$, and cell area, $A_i(t)$, and is the sum of a short-range repulsive term, $U_\text{repel}(\mathbf{r}_{ij}, A_i) = U_0\exp{[-(\mathbf{r}_{ij}/a(A_i))^2]}$ and a harmonic attractive term $U_\text{attract}(\mathbf{r}_{ij}, A_i) = U_1  H(\mathbf{r}_{ij} - a(A_i) - r_a) (\mathbf{r}_{ij} - a(A_i) - r_a)^2$ where $a(A_i) = \sqrt{A_i/\pi}$ is the area-dependent length scale, $H(\cdot)$ is the Heaviside step function, $U_0$ and  $U_1$ are repulsion and harmonic constant respectively, and $r_a$ is a characteristic length scale that quantifies the separation between the repulsive and attractive interaction ranges. 
The stochastic noise $\pmb\eta_i$ is modeled as an Ornstein-Uhlenbeck process with correlation time $\tau$, i.e., $\tau\,\dot{{\pmb\eta}}_i = -\pmb\eta_i + \pmb\xi_i$
where $\pmb\xi_i$ is delta-correlated white noise with zero mean and unit variance.
The noise amplitude $\sigma_i$ is density-dependent such that stochastic fluctuations decrease as cells become more densely packed, i.e., $\sigma_i = \sigma_0\max(2-\rho_i/\rho_0, 0)$, where $\sigma_0$ is a characteristic noise parameter, $\rho_i = 1/A_i$, and $\rho_0$ is a characteristic density parameter. The gradient vector $\mathbf{g}_i$ is defined as the normalized sum of vectors from the cell's centroid to associated front ghost particles (Fig.~\ref{fig:1}{\it D}) and provides a directional bias in the motility toward available free space. The simulation workflow is summarized in Fig.~\ref{fig:1}{\it E}. 

\subsection{Boundary handling and initialization}
Voronoi tessellation offers a natural approach to constructing confluent cell sheets from particle-based models; however, handling unbounded vertices at the system edge is non-trivial. We addressed this by extending the convex hull method of Bobach et al. \cite{bobach2009natural} with an alpha shape construction (Fig.~\ref{fig:1}{\it C}). The alpha shape, a generalization of the convex hull \cite{edelsbrunner2003shape}, captures both convex and concave features of a point set, enabling accurate representation of irregular boundaries (see {\it SI appendix}, Movie~S1).

Simulations were initialized with $N=1520$ particle in a rectangular domain with periodic boundary conditions along the $y$-axis, corresponding to the middle strip of the cell monolayer rather than the full monolayer shown in Fig.~\ref{fig:1}{\it A} and {\it SI appendix}, Fig.~S1. This setup substantially reduces the computational cost compared with simulating the migration of an entire confluent sheet. At each timestep, the alpha shape of the particle set was computed and offset outward by $r_\text{offset} = 0.04$ unit length. Ghost particles were placed along this offset curve and included in the Voronoi tessellation to ensure closure of boundary cells (Fig.~\ref{fig:1}{\it C} and {\it D}). Voronoi polygons intersecting the $y$-domain limits were clipped to remain inside the system. Ghost particles were regenerated at each timestep and interacted only geometrically, thereby stabilizing the boundary without the need to prescribe physical interactions with other particles.

\subsection{Cell division}
Division conditions were evaluated after each SDE update. To prevent synchronous bursts of division, we assigned each particle an initial division time $t_\text{div}$ drawn from a uniform distribution. A cell divided at $t_\text{div}$ only if it satisfied the following three conditions:
\begin{enumerate}
    \item Area threshold: $A_i > A_\text{div} = R_\text{d}^2 = 27.04 \mu$m$^2$(default $R_d = 5.2\mu$m), reflecting the observation that smaller cells generally require a longer growth period before division, while larger cells are more likely to divide \cite{streichan2014spatial, osella2014concerted, zhang2024nuclear}.
    \item Velocity threshold: $|\mathbf{v}_i| < v_\text{max} = 2\mu\text{m hr}^{-1}$, reflecting reduced division in highly motile leader cells \cite{poujade2007collective, park2017tissue}.
    \item Shape constraint: $\max(|\mathbf{r}_i|) > R_0 = 2\mu\text{m}$, where $\max(|\mathbf{r}_i|)$ is the distance between the further vertex and the particle's center of mass, reflecting division suppression under strong confinement \cite{devany2023epithelial}. This constraint is also implemented as a safeguard to numerical instability that may arise due to the new particle being placed too close to the particle it divides from. 
\end{enumerate}
If these conditions were met, the line between the cell particle and its furthest Voronoi vertex is identified, and a new particle is placed at the line's midpoint, akin to the long-axis division observed in experiment \cite{wyatt2015emergence}. The two resulting particles each inherit half of the initial particle’s velocity, their motility vectors are reset to zero, and new division times are drawn from an exponential distribution. Together with the area-dependent interaction term, our division rule provides a simple mechanism for introducing size homeostasis into the model. 

\subsection{Numerical integration}
The coupled SDEs were solved using the Forward-Euler scheme with timestep $\Delta t = 0.01$hr. Each simulation was integrated for $10^4$ steps, corresponding to 100 hours in real life. At each step, Voronoi tessellation and alpha shape reconstruction were updated to compute cell morphologies and topologies.
Fig.~\ref{fig:1}{\it E} provides an overview of the full simulation workflow. A complete list of parameters is provided in {\it SI Appendix} Table~S1 and was used in all simulations unless otherwise specified.

\subsection{Boundary forces}
To prevent penetration through the wall at $x=0$, we applied a harmonic repulsion:
\begin{equation}
	U_x(x) = \begin{cases} k(x_0 - x)^2 &, x \leq x_0 = 2 \mu\text{m} \\
		0&  , x > x_0 = 2 \mu\text{m}\end{cases}
\end{equation}
and damped motility near the wall:
\begin{equation}
    	p_{x, new}(x) = p_x\left(1 - e^{-|x|}\right) , x \leq 2 \mu\text{m},
\end{equation}
where $p_x$ is the $x$-component of the motility vector.

\subsection{Different kind of averaging}
Statistical analyses are performed using ensemble-averaged quantities. We use an overline to denote averaging within a single simulation at a given time $t$. For instance, the mean area of all cells in the simulation at a particular time is
\begin{equation}
\overline{A} = \frac{1}{N}\sum_{j=1}^N A_j,
\end{equation} 
where $N$ is the total number of cells at time $t$. 

To compute the local mean cell area, $A(x)$, particles are coarse-grained into ten spatial bins according to their normalized $x$-position from the end, with $x=0$ corresponding to the first bin and $x=1.0$ to the last bin. Each bin spans the full range of $y$. Let $B(x)$ denote the set of particles in the $x$-th bin and $N(x)$ be the number of particles in the $x$-th bin. The local mean cell area in the $x$-th bin, $A(x)$, is defined as 
\begin{equation}
   A(x) = \frac{1}{N(x)}\sum_{i \in B(x)} A_i.
\end{equation}
The curves shown in Fig.~2{\it C}-{\it E} are obtained via piecewise interpolation of these ten points, $A(x)$. 

We use the notation $\langle \cdot \rangle$ to denote ensemble averaging over $10$ independent simulations. In particular, the ensemble average of the mean cell area $\langle \overline{A} \rangle$ is defined as
\begin{equation}
\langle \overline{A} \rangle = \frac{1}{10}\sum_{j=1}^{10}\frac{1}{N^{(j)}}\sum_i^{N^{(j)}} A_{i}^{(j)} = \frac{1}{10}\sum_{j=1}^{10}\ \overline{A^{(j)}},
\end{equation} 
where $A_{i}^{(j)}$ is the area of the $i$-th cell in the $j$-th simulation, $N^{(j)}$ is the total number of particles in the $j$-th simulation, and $\overline{A^{(j)}}$ is the mean cell area in the $j$-th simulation. Similar definitions apply for the ensemble-averaged local mean cell area, $\langle A(x)\rangle$, ensemble-averaged mean speed, $\langle \overline{v}\rangle$, and ensemble-averaged local mean hexatic order parameter $\langle {|\gamma_6(x)|} \rangle$ among others.

\section{Results}

\subsection{Hexanematic order arising from growing boundary}
\begin{figure*}[!t]
    \centering
	\includegraphics[width=\linewidth]{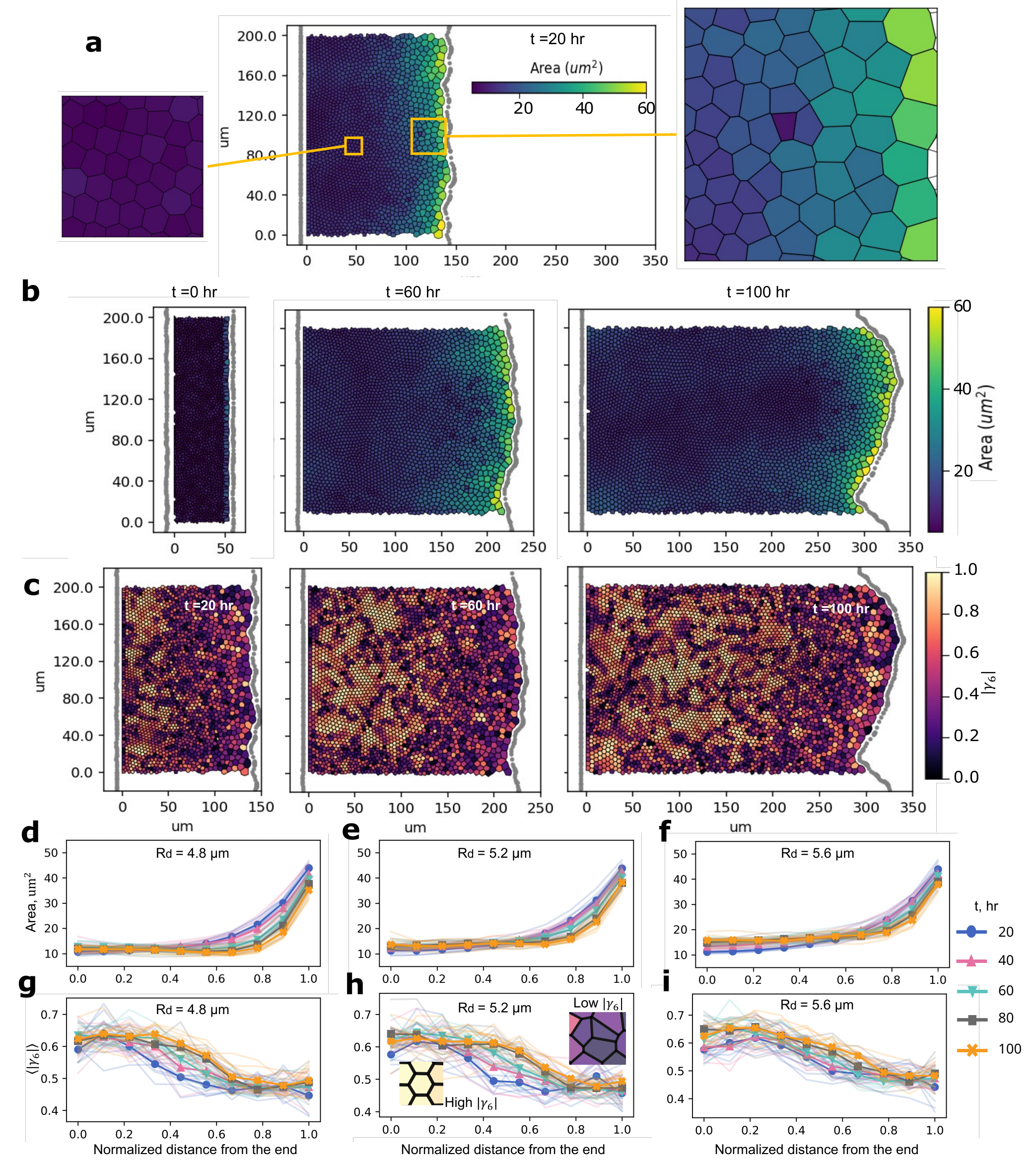}
	\caption{Cell morphology in simulation. 
	({\it A}) Representative Voronoi tessellation at $t = 20$ hr for $R_{d} = 5.2\mu$m and $\beta = 0.12$. Insets highlight peripheral and bulk regions: bulk cells are smaller and more regularly packed, while peripheral cells are larger and more irregular. 
({\it B} and {\it C}) Representative tessellations at $t = 20, 60,$ and $100$ hr showing evolution of cell area and hexatic order over time. 
({\it D-F}) Ensemble-averaged local mean cell area, $\langle A(x)\rangle$, and ({\it G-I}) ensemble-averaged local mean hexatic order, $\langle |\gamma_6(x)|\rangle$ as functions of normalized end-to-end distance along cell sheet at five time points ($t = 20, 40, 60, 80, 100$ hr) for $\beta = 0.16$ and three different minimum division lengths $R_\text{d}$. 
Dark lines show ensemble averages over 10 simulations; light lines correspond to individual runs. 
Insets in ({\it H}) show typical cell morphologies with high and low $|\gamma_6|$.}
\label{fig:2}
\end{figure*}

We begin by examining the evolution of cell morphology predicted by our stochastic particle--Voronoi model. The simulations are initialized with nearly uniform cell sizes and no discernible morphological organization, with an ensemble-averaged mean area at $t=0$ of $\langle \bar{A} \rangle_{t=0} \approx 7.1\mu\text{m}^2$. We use $\langle \cdot \rangle$ to denote ensemble-averaging. Upon release, the initially quiescent monolayer migrates collectively toward the empty region on the right ($+x$ direction) (Fig.~\ref{fig:2}{\it A} and {\it B}). We acknowledge that this initial density is higher than many experimentally observed values. However, this discrepancy is expected to primarily affect early transient dynamics, since the ensemble-averaged mean speed reaches a steady state by $t\approx20$ hr (Fig.~\ref{fig:3}{\it D}), consistent with the $t\geq20$ window that we will be using throughout our analysis. We additionally tested our model at a lower initial density and observed a consistent morphology trend (discussed further in the following paragraphs) despite the reduced initial density. We invite curious readers to see {\it SI Appendix} for further discussion of the effect of initial density on our findings.

Although the monolayer is initially homogeneous, a clear front-bulk segregation develops quickly over time. Voronoi tessellations at $t=20,60,100$ hr reveal the spontaneous emergence of pronounced spatial heterogeneity. Cells at the advancing boundary progressively increase in area and become more irregular, deviating from hexagonal packing, whereas bulk cells remain smaller, more densely packed, and predominantly hexagonal (Fig.~\ref{fig:2}{\it A}, {\it B} and {\it SI appendix}, Movie~S1). This spatial differentiation captures both the enlargement of peripheral cells and the concomitant loss of hexagonal order near the migrating edge seen in Figure \ref{fig:1}A (see {\it SI Appendix} SFig 1 and 2 for additional analysis on the morphology)\cite{matsuzawa2018alpha}.

At each time point, we compute the local mean cell area, $A(x)$, as a function of the normalized position, $x$, along the confluent sheet with $x=0$ at the rear and $x=1$ at the front. The resulting spatial profile exhibits a clear monotonic increase in average area from the rear toward the advancing front. While the ensemble-averaged local mean cell area $\langle A(x)\rangle$ remains approximately constant throughout the bulk region, it increases sharply beyond ~60\% of the sheet length measured from the rear, marking the onset of a boundary layer (see {\it Methods}). Notably, cells at the periphery can attain areas up to fivefold larger than those in the bulk, underscoring the pronounced front-bulk morphological segregation. This fold-change is in qualitative agreement with our simple image analysis of the experimental HeLa monolayers. Remarkably, varying division thresholds, $R_\text{d}$, does not significantly alter the spatial profile of the local mean cell area (Fig.~\ref{fig:2}{\it D}-{\it F}), indicating that the area gradient is robust to changes in proliferation rate. This insensitivity arises from a compensating mechanism, where more frequent division at low $R_\text{d}$ produces smaller average cell sizes, increasing local packing density and mechanical drag, which offsets the additional proliferative flux. In this sense, cell migration is governed primarily by mechanical relaxation and boundary-driven crawling rather than by the net growth rate alone, a robustness that is dynamical in origin.

To quantify cell shape anisotropy, we compute the hexatic order parameter, $|\gamma_6|$, for each cell \cite{armengol2023epithelia, armengol2024hydrodynamics}, defined as
$\gamma_6 = \Delta^{-1} \sum_{i}|\mathbf{r}_i|^6 e^{6i\phi_i}$ where the sum runs all the vertices of the polygonal contour of a cell, $\phi_i$ is the corresponding polar angle, and $\Delta = \sum_{i} |\mathbf{r}_i|^6$,  $\mathbf{r}_i$ denotes the displacement from the cell centroid to vertex $i$. The magnitude of $|\gamma_6|$ measures the degree of sixfold symmetry: values close to unity correspond to nearly hexagonal cells, whereas smaller values indicate increasing distortion, elongation, or irregularity.

At each time point, we compute the local mean hexatic order parameter, $|\gamma_6(x)|$, as a function of the normalized position, $x$, along the confluent sheet with $x=0$ at the rear and $x=1$ at the front. The spatial profile of the local mean hexatic order parameter exhibits a trend opposite to that of the mean cell area. The highest values of the ensemble-averaged local mean hexatic order parameter $\langle |\gamma_6(x)|\rangle$ are found in the interior of the sheet (normalized distance $x\sim 0.1$-$0.3$), reflecting strong hexagonal packing in the bulk. Moving toward the advancing front, $\langle |\gamma_6(x)|\rangle$ decreases progressively, reaching its minimum near the boundary (Fig.~\ref{fig:2}{\it C} and {\it SI appendix}, Movie~S2). This gradual decline indicates a continuous breakdown of sixfold orientational order toward the leading edge.

Importantly, the reduction in $\langle |\gamma_6(x)|\rangle$ extends over multiple cell layers rather than being confined to the outermost boundary row, demonstrating that the loss of hexagonal order represents a finite-width structural transition zone rather than a trivial boundary artifact. This spatial trend persists across all examined time points ($t=20, 40, 60, 80, 100$ hr) and remains robust under variations in the division threshold $R_\text{d}$ (Fig.~\ref{fig:2}{\it G}-{\it I}). Together with the corresponding increase in mean cell area near the front, these results provide strong evidence for the emergence of hexanematic organization across the monolayer. The inverse correlation between hexatic order and cell area is important for the emergence of hexanematic separations - as cells migrate into free space, local packing constraints are progressively released, allowing cells to spread and elongate. Conversely, cells in the interior remain densely confined by their neighbors, preserving hexagonal packing and high hexatic order. This competition naturally gives rise to the observed hexanematic organization.

\subsection{Cell migration driven by boundary dynamics}

\begin{figure*}[htb!]
    \centering
	\includegraphics[width=\textwidth]{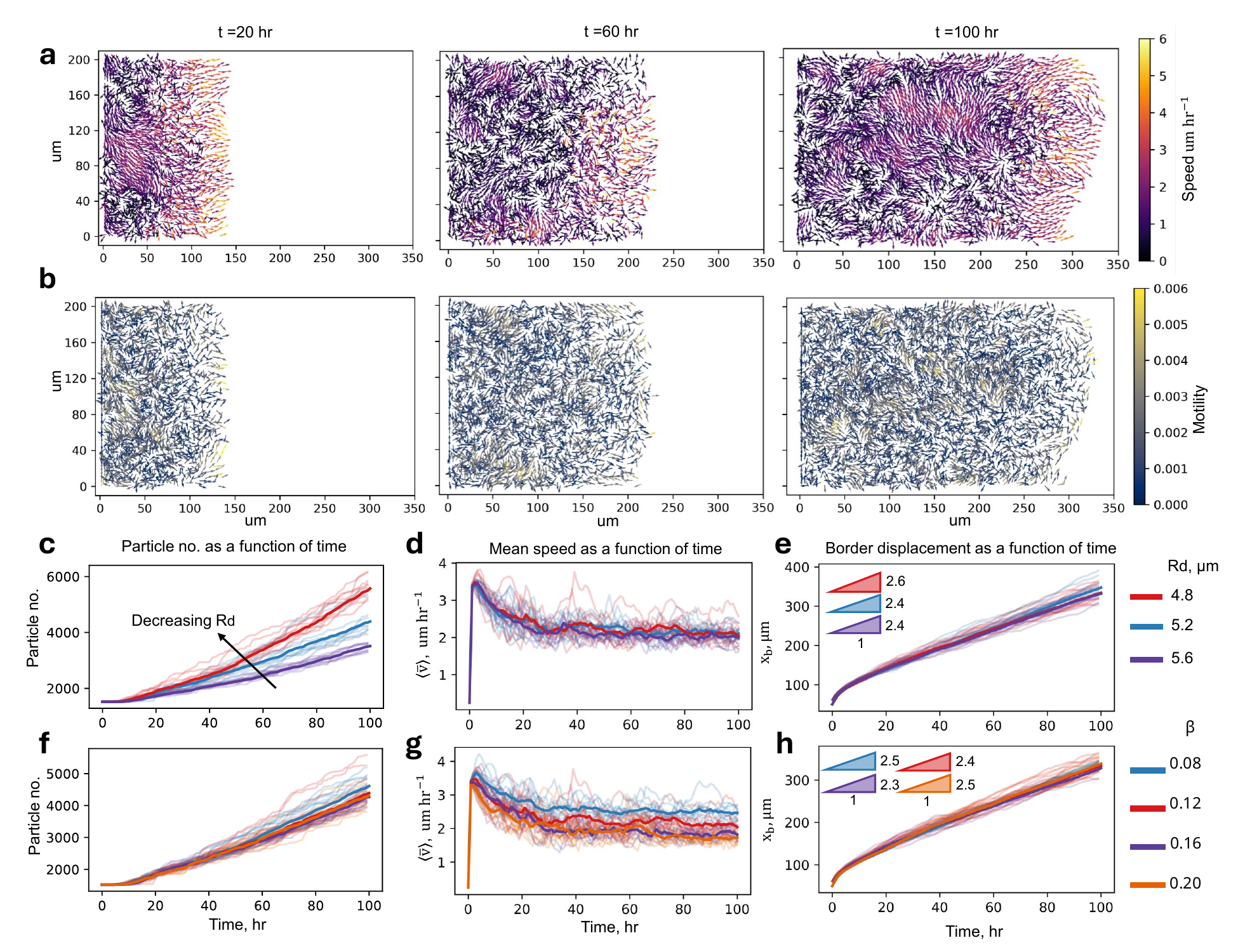}
	\caption{Dynamics of cell migration.
	({\it A} and {\it B}) Representative velocity and motility fields at $t = 20, 60,$ and $100$ hr. Arrow color encodes magnitude (yellow = larger). Velocities are generally higher at the periphery than in the bulk.
	({\it C} and {\it F}) Total particle number, $N_t$, ({\it D} and {\it G} mean speed $\bar{v}$, and ({\it E} and {\it H}) maximum border displacement, $x_b$, as a function of time for different minimum division lengths, $R_\text{d}$, and motility coupling constants, $\beta$. Bold lines denote ensemble averaging across 10 independent simulations; fine lines (in lighter colors) show individual runs. 
	}
	\label{fig:3}
\end{figure*}

 The velocity fields, $\mathbf{v}_i(t)$, (Fig.~\ref{fig:3}{\it A}) show that peripheral cells move towards the free space (on the right) and are significantly faster than bulk cells, in agreement with experimental observations \cite{okimura2022leading, petitjean2010velocity} and the Sep{\'u}lveda model \cite{sepulveda2013collective}. In contrast, despite the addition of gradient vector to the cell at the boundary, the motility fields, $\mathbf{m}_i(t)$, (Fig.~\ref{fig:3}{\it B}) do not display a clear front-bulk separation, i.e., cells with larger motility vectors appear not only at the peripheral but throughout the monolayer, indicating that the observed velocity heterogeneity arises primarily from collective mechanical interactions rather than from an imposed spatial variation in intrinsic motility. 

 In our framework, the asymmetric boundary (open space ahead, confined tissue behind) permits peripheral cells to relax outward under repulsive interactions alone, reducing local density and promoting larger cell areas at the periphery even without active motility. To test this directly, we repeated our simulations with the velocity–motility coupling set to zero ($\kappa = 0$, see {\it SI Appendix} SFig 5). The resulting spatial profiles of cell area and hexatic order closely resemble those obtained with motility intact, confirming that the segregation pattern is substantially governed by repulsive relaxation at the free boundary rather than requiring active self-propulsion. Bulk cell migration, however, is essentially abolished under these conditions. Peripheral cells continue to advance into free space under repulsion alone, while bulk cells, symmetrically confined by neighbors, show minimal net motion without active motility. Recall our division rule, where cells exceeding a prescribed velocity threshold, or too confined to satisfy the area and shape constraint, are inhibited from dividing. The division rule acts together with the kinetic asymmetry established by the boundary to reinforce the density gradient, where the suppressed division at the periphery and in the confined bulk alike prevents replenishment that would otherwise relax the gradient, thus stabilizing the segregation pattern over time without being responsible for its initial emergence

Next, we investigate the collective dynamics by tracking the total particle number, $N_{t}$, the mean speed, $\overline{v}$, and the maximum border displacement, $x_b$, under varying division thresholds, $R_\text{d}$, and motility coupling strengths, $\beta$. As expected, particle number increases more rapidly for smaller $R_\text{d}$, reflecting higher rate of cell divisions (Fig.~\ref{fig:3}{\it C}). The ensemble-averaged mean speed, $\langle \overline{v} \rangle$, rises sharply during the initial equilibration, reaches a maximum, and subsequently decays towards a steady value of $2-3 \mu$m hr$^{-1}$ after $\sim$20 hr (Fig.~\ref{fig:3}{\it D}). The non-monotonic transient behavior in ensemble-averaged mean speed is qualitatively consistent with experimental reports in MDCK monolayers, where cell speed peaks immediately after stencil removal and relaxes as the tissue reorganizes \cite{petitjean2010velocity}. Mechanistically, the initial velocity surge reflects the rapid release of mechanical confinement and the establishment of outward flux, whereas the subsequent decay corresponds to the progressive buildup of intercellular constraints and collective stress within the expanding sheet. 

Interestingly, although reducing $R_\text{d}$ increases the division rate, it has little effect on the mean speed. A similar insensitivity is observed in the border displacement (Fig.~\ref{fig:3}{\it E}), which varies only modestly across division thresholds. A plausible explanation is that increased division reduces the average cell size, thereby producing a denser monolayer. The resulting crowding effect mitigates the impact of proliferation on collective progression. In this sense, cell migration is also co-governed by boundary-driven crawling behavior rather than by the net proliferation rate or mechanical relaxation only.

\begin{figure*}[hbt!]
    \centering
\includegraphics[width=\textwidth]{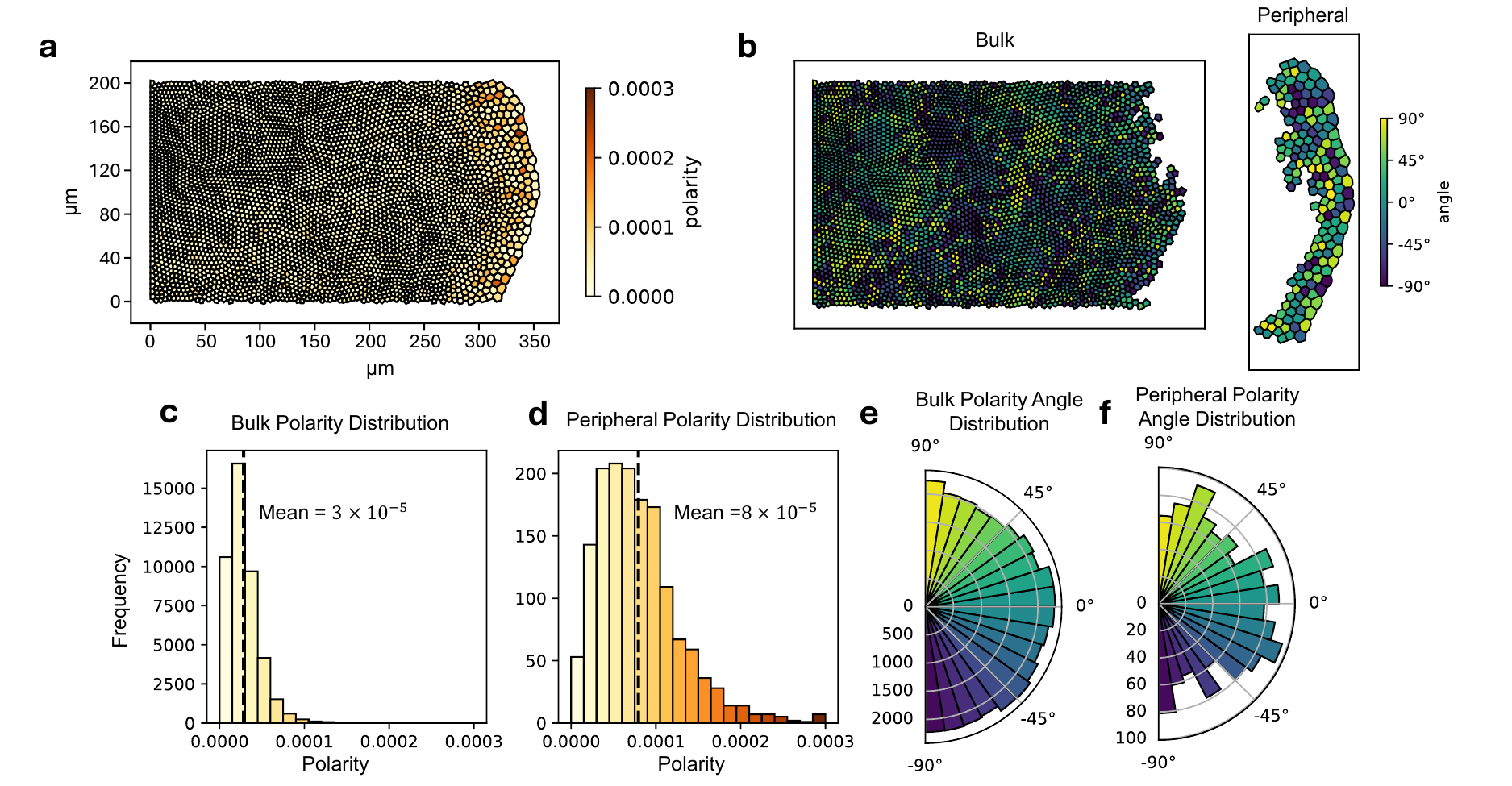}
\caption{Cell polarity statistics. 
({\it A} and {\it B}) Representative cell polarity map at $t = 100$ hr for $\beta = 0.16$ and $R_\text{d} = 5.2\mu$m. The yellow-brown colormap ({\it A}) indicates polarity magnitude, while the blue-yellow colormap ({\it B}) encodes the angle of the largest eigenvector relative to the $x$-axis. 
({\it C} and {\it D}) Histogram of cell polarity magnitude and ({\it E} and {\it F}) polar plot of cell polarity orientation, averaged over 10 independent simulations. 
}
\label{fig:4}
\end{figure*}

By contrast, varying $\beta$ has little effect on population size but strongly reduces ensemble-averaged speed (Fig.~\ref{fig:3}{\it F} and {\it G}). Increasing $\beta$ strengthens the alignment (or coupling) between neighboring cell motility, effectively enhancing collective coherence while suppressing individual fluctuations.  Despite this pronounced reduction in bulk speed, the maximal border displacement remains largely insensitive to $\beta$ (Fig.~\ref{fig:3}{\it H}). This indicates that motility coupling primarily reorganizes the internal velocity field, i.e., modifying spatial correlations and flow structures, without significantly affecting the speed in which tissue boundary progresses. This probably increased coordination among peripheral cells at higher $\beta$, compensating for the drop in mean speed and allowing cells at the front to move in a more coordinated, cooperative manner even as individual speeds decline. The maintenance of border displacement speed despite the drop in mean single cell speed as $\beta$ increases, together with the ablation test when $\kappa = 0$, suggests that motility governs the dynamics layered on top of the mechanical relaxation that arises from an asymmetric tissue boundary. Higher motility alignment between neighboring cells $\beta$ promotes coordinated, cooperative motion and enhances the coherence of migration, without which cells in the interior of the tissue would remain largely immobile, and the migration would be purely repulsion-driven.

The gradient of the border displacement becomes approximately linear after $t \approx 10$ hr under all conditions with average border speed of approximately $2.5\mu$m/hr (Fig.~\ref{fig:3}{\it E} and {\it H}),lower than the experimentally measured mean border speed of 6.0$\mu$m/hr across 11 independent experiments (Fig.~\ref{fig:1}{\it B}).

\subsection{Peripheral cells have higher polarity than bulk cells} 
We further quantified the local morphology by studying the cell elongation, i.e., polarity, $\mathbf{p}_i$, using principal component analysis (PCA) of the Voronoi vertex positions. In this framework, the covariance matrix of vertex coordinates defines the cell shape tensor. Its eigenvectors, $\mathbf{e}_1$ and $\mathbf{e}_2$, determine the principal axes of the cell, while the corresponding eigenvalues, $\lambda_1$ and $\lambda_2$, quantify the variance along those axes. For isotropic cells (i.e., hexatic), the eigenvalues are nearly equal. In contrast, elongated cells (i.e., nematic), exhibit a pronounced spectral gap, with one dominant eigenvalue, $\lambda_1$. We therefore define the polarity magnitude, $|\mathbf{p}_i|$, as the difference between the two eigenvalues, $\lambda_1-\lambda_2$, and the polarity director, $\hat{\mathbf{p}}_i = \hat{\mathbf{e}}_1$, as the normalized eigenvector of the larger eigenvalue, thereby capturing both the strength and orientation of elongation.

A representative polarity map at $t = 100$ hr for $\beta = 0.16$ and $R_\text{d} = 5.2 \mu$m is illustrated in Fig.~\ref{fig:4}{\it A} and {\it B}. Cells at the moving front are classified as peripheral or bulk according to their area, with area $A \geq 27.8\mu\text{m}^2$ classified as peripheral and $A < 27.8\mu\text{m}^2$ as bulk (see {\it Methods}). The spatial map reveals that elongated cells (i.e., nematic) are predominantly localized near the advancing front, whereas bulk cells remain comparatively isotropic (i.e., hexatic) and smaller in size.

This distinction is quantified in the polarity magnitude distributions (Fig.~\ref{fig:4}{\it C} and {\it D}). Peripheral cells exhibit a clear shift toward larger polarity values, with a higher mean and broader distribution relative to bulk cells. This result is consistent with experimental observations that cells near the wound edge become more elongated and anisotropic during collective migration \cite{reffay2011orientation}.

We further examined polarity orientation by computing the angle of each polarity vector with respect to the $x$-axis (the migration direction). We treat the polarity director as a nematic vector, i.e., $\hat{\mathbf{p}}_i  \equiv -\hat{\mathbf{p}}_i$, and thus the angles map onto $[-90^{\circ}, 90^{\circ}]$. The polar histograms (Fig.~\ref{fig:4}{\it E} and {\it F}) reveal that bulk cells display a near uniform angular distribution while peripheral cells are slightly peaked in the forward direction, reflecting reduced packing constraints and local rearrangements at the active boundary.

\begin{figure*}[htb!]
    \centering
	\includegraphics[width =\textwidth]{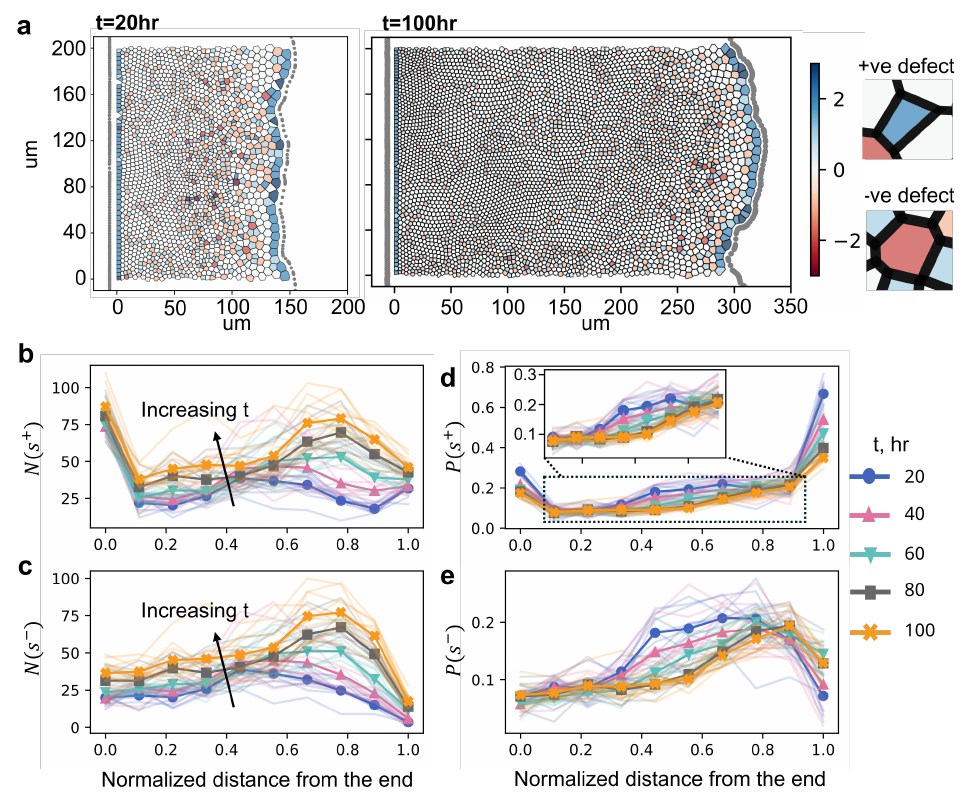}
    \caption{Topological defects in cell migration. 
    ({\it A}) Representative Voronoi tessellation at $t = 20$ hr and $t = 100$ hr for $R_{d} = 5.6\mu$m and $\beta = 0.12$. Blue/red cells denote positive/negative topological defects as illustrated schematically on the right.
    ({\it B}) and ({\it C}) Spatial distribution of ({\it B}) positive and ({\it C}) negative topological defect, $N^+(x)$ and $N^-(x)$, as functions of the normalized distance from the tissue edge at $t = 20, 40, 60, 80$, and $100$ hr. 
    ({\it D}) and ({\it E}) Probability of ({\it D}) positive and ({\it E}) negative topological defect as functions of normalized end-to-end distance along cell sheet at $t = 20, 40, 60, 80$, and $100$ hr. The inset in ({\it D}) shows the same plot but excluding the first and last points for clarity. The probability of positive and negative topological defects in general increases by 2 fold at the periphery compared to the bulk. 
        }
    \label{fig:5}
\end{figure*}

\subsection*{Peripheral cells are more defect-prone} 
We quantify topological defects by computing the coordination number $z$ of each cell, defined as the number of neighboring cells that share an edge. In a two-dimensional Voronoi tessellation, cells typically exhibit sixfold coordination. The topological defect number is defined as $s=6-z$, so that any deviation from sixfold coordination reflects local structural disorder \cite{lidmar2003virus, yong2013elastic}. Cells with $z < 6$ correspond to positive topological defects (positive disclinations), whereas cells with $z > 6$ correspond to negative defects (see Fig.~\ref{fig:5}).

By construction, cells at the outermost and innermost layers exhibit positive defects due to missing neighbors at the boundaries. Although we include these cells in all plots, we focus our analysis on cells away from the boundary layers to minimize boundary-induced artifacts. We also note that the defect number $s$ is commonly interpreted as the topological charge in Voronoi-based models \cite{jimenez2016curvature}.

As the tissue grows, the total number of positive and negative topological defects increases moderately with time (see Fig.~\ref{fig:5}{\it B}, {\it C}, and {\it SI appendix}, SFig.6). 
More strikingly, the defects exhibit a robust spatial organization: their density increases gradually from the rear toward the front, reaches a maximum in the near-front region, and then declines near the leading edge. Because particle density decreases toward the front as cell size increases, we normalize for this effect by computing the probability distributions of positive and negative defects as functions of the normalized position $x$:
$P(s^+) =N^+(x)/N(x)$ and $P(s^-)=N^-(x)/N(x)$, 
where $N^\pm(x)$ and $N(x)$ denote the number of positive/negative defects and the total number of cells in region $x$, respectively (see Fig.~\ref{fig:5}{\it D}-{\it E}).

These normalized profiles show that the periphery is significantly more defect-prone than the bulk, with defect fractions increasing from $\sim 10\%$ in the bulk to $\sim 20\%$ near the front. This enrichment indicates enhanced structural disorder and active rearrangements in the advancing region. Mechanistically, the accumulation of defects at the periphery likely arises from the combined effects of elevated cell velocities and reduced local packing density, both of which promote local distortions of hexatic order and facilitate topological defect formation.

\section{Discussion and Conclusion}
In this study, we developed a stochastic particle model of collective migration in confluent epithelial sheets, incorporating Voronoi tessellation and an alpha shape extension to handle tissue boundaries. This framework allows us to capture both cell morphology and neighbor interactions while maintaining flexibility to introduce additional mechanical processes, including motility alignment and density-dependent responses. 

We showed that hexanematic separation is an emergent, mechanically driven phenomenon arising from boundary-induced collective dynamics, not a consequence of imposed heterogeneity. This conclusion is supported directly by the dissociation between the velocity and motility fields: peripheral cells move systematically faster than bulk cells, yet their intrinsic motility vectors show no comparable front-bulk separation, indicating that the observed heterogeneity is generated by collective mechanical interactions rather than by any prescribed spatial variation in single-cell behavior.

Mechanistically, active boundary establishes an outward velocity gradient across the sheet, which drives hexanematic separation through two coupled effects: first, the resulting reduction in local density near the advancing front mechanically relaxes packing constraints, permitting peripheral cells to spread and elongate while bulk cells remain confined and hexagonally packed; second, our division rule directly links this spatial pattern to proliferation through two complementary gates — a velocity threshold that suppresses division in fast-moving peripheral cells, and both the area and shape constraints that suppress division in tightly confined, geometrically isotropic cells. Together, these mechanisms convert a purely kinematic asymmetry, faster motion at the front due to the active boundary, into a stable, self-sustaining pattern of morphological and orientational order, without requiring distinct cell types, prescribed target shapes, or any built-in spatial heterogeneity.

Despite these successes, the model has limitations. First, our division rule predicts a local peak in division specifically at the interface between periphery and bulk. While this detailed spatial prediction has not, to the best of our knowledge, been directly tested experimentally, some indirect evidence is consistent with it. Zhang et al. performed single-cell elongation experiments on HeLa cells and found that intermediate cell length is optimal for division and proliferation, with reduced division at both shorter and longer lengths \cite{zhang2024nuclear}. This non-monotonic dependence qualitatively resembles our model's prediction, in which cells at the periphery-bulk interface, which are intermediate in size between the small, confined bulk cells and the large, fast peripheral cells, divide more readily than cells at either extreme. We note, however, that Zhang et al.'s experiment probes single-cell elongation directly, so this correspondence should be regarded as suggestive rather than a direct experimental validation of the specific mechanism proposed here. It would be of interest to test experimentally whether cells at the periphery-bulk interface in a migrating monolayer indeed exhibit a higher division rate than cells at either extreme.

Second, although it produces a coherent advancing front, the digitated, finger-like protrusions reported by Sep{\'u}lveda et al. \cite{sepulveda2013collective} are less pronounced in our model. In their framework, these structures emerge from the inclusion of ``leader” particles with distinct behavior and surface particles that can be randomly removed, mimicking local breaks in surface tension. As a direction for future work, we propose incorporating an energy term describing the mechanics of the cell membrane into the particle dynamics, similar to the VM and SPV models. Such an extension could allow the framework to capture finer morphological features and more realistic tissue dynamics. In particular, this framework may provide a promising platform to explore the jamming-unjamming transition within a single confluent sheet directly, avoiding the need to artificially separate jammed and unjammed phases into different simulations. In addition, a more careful modeling of motility - one that describes a more realistic chemical and mechanical signals transmission within and between cells - may further improve the model’s predictive accuracy and potentially reproduce finer morphologies such as digitated protrusions. 

Overall, our model provides a minimal yet mechanistically grounded description of how the hexanematic order may emerge purely from mechanically mediated interactions, differential cell growth and division, and an explicitly active tissue boundary that biases motion toward the front. By bridging the flexibility of particle-based approaches with the geometric resolution of vertex and SPV models, this framework provides a versatile platform for investigating the coupled evolution of mechanics, morphology, and collective tissue dynamics.

\section*{Author contributions}

H.L.T. and E.H.Y. designed research; 
D.Z. performed the experiment; 
F.D. and D.E.T. performed initial computational studies; 
H.L.T. and I.S.Y.L. performed computational studies; 
H.L.T., H.L., and E.H.Y. analyzed data; 
E.H.Y. coordinated and supervised the research; 
H.L.T. wrote the initial draft; 
H.L.T., H.L., and E.H.Y. contributed to the writing and review of the manuscript.

\section*{Conflicts of interest}
There are no conflicts to declare.

\section*{Data availability}
All data supporting this study are included in the article and/or the SI Appendix. The simulation code is available at \href{https://github.com/honlin96/Stochastic-Particle-Voronoi-Cell-Model-}{https://github.com/honlin96/Stochastic-Particle-
Voronoi-Cell-Model-}.

\section*{Acknowledgements}

H.L.T., F. D., and E.H.Y. acknowledge support from Singapore Ministry of Education through the Academic Research Fund Tier 1 (RG140/22) and Academic Research Fund Tier 2 (MOE-T2EP50223-0014). 
H.L. acknowledges support from the National Key Research and Development Program of China (Grant No. 2025YFA0922902).




\bibliography{references} 

@article{zhang2024nuclear,
  title={Nuclear deformation and cell division of single cell on elongated micropatterned substrates fabricated by DMD lithography},
  author={Zhang, Duo and Wu, Wenjie and Zhang, Wanying and Feng, Qiyu and Zhang, Qingchuan and Liang, Haiyi},
  journal={Biofabrication},
  volume={16},
  number={3},
  pages={035001},
  year={2024},
  publisher={IOP Publishing}
}

@article{campas2009shape,
  title={Shape and dynamics of tip-growing cells},
  author={Campas, Otger and Mahadevan, L},
  journal={Current Biology},
  volume={19},
  number={24},
  pages={2102--2107},
  year={2009},
  publisher={Elsevier}
}

@article{lidmar2003virus,
  title={Virus shapes and buckling transitions in spherical shells},
  author={Lidmar, Jack and Mirny, Leonid and Nelson, David R},
  journal={Physical Review E},
  volume={68},
  number={5},
  pages={051910},
  year={2003},
  publisher={APS}
}

@article{brugues2014forces,
  title={Forces driving epithelial wound healing},
  author={Brugu{\'e}s, Agust{\'\i} and Anon, Ester and Conte, Vito and Veldhuis, Jim H and Gupta, Mukund and Colombelli, Julien and Mu{\~n}oz, Jos{\'e} J and Brodland, G Wayne and Ladoux, Benoit and Trepat, Xavier},
  journal={Nature physics},
  volume={10},
  number={9},
  pages={683--690},
  year={2014},
  publisher={Nature Publishing Group UK London}
}

@article{fenteany2000signaling,
  title={Signaling pathways and cell mechanics involved in wound closure by epithelial cell sheets},
  author={Fenteany, Gabriel and Janmey, Paul A and Stossel, Thomas P},
  journal={Current biology},
  volume={10},
  number={14},
  pages={831--838},
  year={2000},
  publisher={Elsevier}
}

@article{notbohm2016cellular,
  title={Cellular contraction and polarization drive collective cellular motion},
  author={Notbohm, Jacob and Banerjee, Shiladitya and Utuje, Kazage JC and Gweon, Bomi and Jang, Hwanseok and Park, Yongdoo and Shin, Jennifer and Butler, James P and Fredberg, Jeffrey J and Marchetti, M Cristina},
  journal={Biophysical journal},
  volume={110},
  number={12},
  pages={2729--2738},
  year={2016},
  publisher={Elsevier}
}

@article{alert2020physical,
  title={Physical models of collective cell migration},
  author={Alert, Ricard and Trepat, Xavier},
  journal={Annual Review of Condensed Matter Physics},
  volume={11},
  number={1},
  pages={77--101},
  year={2020},
  publisher={Annual Reviews}
}

@article{vicsek2012collective,
  title={Collective motion},
  author={Vicsek, Tam{\'a}s and Zafeiris, Anna},
  journal={Physics reports},
  volume={517},
  number={3-4},
  pages={71--140},
  year={2012},
  publisher={Elsevier}
}

@article{matsuzawa2018alpha,
	title={$\alpha$-Catenin controls the anisotropy of force distribution at cell-cell junctions during collective cell migration},
	author={Matsuzawa, Kenji and Himoto, Takuya and Mochizuki, Yuki and Ikenouchi, Junichi},
	journal={Cell Reports},
	volume={23},
	number={12},
	pages={3447--3456},
	year={2018},
	publisher={Elsevier},
	doi = {10.1016/j.celrep.2018.05.070}
}

@article{pasupalak2020hexatic,
  title={Hexatic phase in a model of active biological tissues},
  author={Pasupalak, Anshuman and Yan-Wei, Li and Ni, Ran and Ciamarra, Massimo Pica},
  journal={Soft matter},
  volume={16},
  number={16},
  pages={3914--3920},
  year={2020},
  publisher={Royal Society of Chemistry}
}

@article{lander2011pattern,
  title={Pattern, growth, and control},
  author={Lander, Arthur D},
  journal={Cell},
  volume={144},
  number={6},
  pages={955--969},
  year={2011},
  publisher={Elsevier}
}

@article{manukyan2017living,
  title={A living mesoscopic cellular automaton made of skin scales},
  author={Manukyan, Liana and Montandon, Sophie A and Fofonjka, Anamarija and Smirnov, Stanislav and Milinkovitch, Michel C},
  journal={Nature},
  volume={544},
  number={7649},
  pages={173--179},
  year={2017},
  publisher={Nature Publishing Group UK London}
}

@article{okimura2022leading,
  title={Leading-edge elongation by follower cell interruption in advancing epithelial cell sheets},
  author={Okimura, Chika and Iwanaga, Misaki and Sakurai, Tatsunari and Ueno, Tasuku and Urano, Yasuteru and Iwadate, Yoshiaki},
  journal={Proceedings of the National Academy of Sciences},
  volume={119},
  number={18},
  pages={e2119903119},
  year={2022},
  publisher={National Academy of Sciences}
}

@article{wyatt2015emergence,
	title={Emergence of homeostatic epithelial packing and stress dissipation through divisions oriented along the long cell axis},
	author={Wyatt, Tom PJ and Harris, Andrew R and Lam, Maxine and Cheng, Qian and Bellis, Julien and Dimitracopoulos, Andrea and Kabla, Alexandre J and Charras, Guillaume T and Baum, Buzz},
	journal={Proceedings of the National Academy of Sciences},
	volume={112},
	number={18},
	pages={5726--5731},
	year={2015},
	publisher={National Academy of Sciences},
	doi = {10.1073/pnas.1420585112}
}

@article{yang2016probing,
  title={Probing leader cells in endothelial collective migration by plasma lithography geometric confinement},
  author={Yang, Yongliang and Jamilpour, Nima and Yao, Baoyin and Dean, Zachary S and Riahi, Reza and Wong, Pak Kin},
  journal={Scientific reports},
  volume={6},
  number={1},
  pages={22707},
  year={2016},
  publisher={Nature Publishing Group UK London}
}

@article{yong2013elastic,
  title={Elastic platonic shells},
  author={Yong, Ee Hou and Nelson, David R and Mahadevan, Lakshminarayanan},
  journal={Physical review letters},
  volume={111},
  number={17},
  pages={177801},
  year={2013},
  publisher={APS}
}

@article{bi2016motility,
  title={Motility-driven glass and jamming transitions in biological tissues},
  author={Bi, Dapeng and Yang, Xingbo and Marchetti, M Cristina and Manning, M Lisa},
  journal={Physical Review X},
  volume={6},
  number={2},
  pages={021011},
  year={2016},
  publisher={APS}
}

@article{poujade2007collective,
  title={Collective migration of an epithelial monolayer in response to a model wound},
  author={Poujade, Mathieu and Grasland-Mongrain, Erwan and Hertzog, Ariane and Jouanneau, Julien and Chavrier, Philippe and Ladoux, Beno{\^\i}t and Buguin, Axel and Silberzan, Pascal},
  journal={Proceedings of the National Academy of Sciences},
  volume={104},
  number={41},
  pages={15988--15993},
  year={2007},
  publisher={National Academy of Sciences},
  doi = {10.1073/pnas.0705062104}
}

@article{streichan2014spatial,
  title={Spatial constraints control cell proliferation in tissues},
  author={Streichan, Sebastian J and Hoerner, Christian R and Schneidt, Tatjana and Holzer, Daniela and Hufnagel, Lars},
  journal={Proceedings of the National Academy of Sciences},
  volume={111},
  number={15},
  pages={5586--5591},
  year={2014},
  publisher={National Academy of Sciences}
}

@article{park2017tissue,
  title={Tissue-scale coordination of cellular behaviour promotes epidermal wound repair in live mice},
  author={Park, Sangbum and Gonzalez, David G and Guirao, Boris and Boucher, Jonathan D and Cockburn, Katie and Marsh, Edward D and Mesa, Kailin R and Brown, Samara and Rompolas, Panteleimon and Haberman, Ann M and others},
  journal={Nature cell biology},
  volume={19},
  number={3},
  pages={155--163},
  year={2017},
  publisher={Nature Publishing Group UK London}
}

@article{edelsbrunner2003shape,
  title={On the shape of a set of points in the plane},
  author={Edelsbrunner, Herbert and Kirkpatrick, David and Seidel, Raimund},
  journal={IEEE Transactions on information theory},
  volume={29},
  number={4},
  pages={551--559},
  year={2003},
  publisher={IEEE},
  doi = {10.1109/TIT.1983.1056714}
}

@article{bobach2009natural,
  title={Natural neighbor extrapolation using ghost points},
  author={Bobach, Tom and Farin, Gerald and Hansford, Dianne and Umlauf, Georg},
  journal={Computer-Aided Design},
  volume={41},
  number={5},
  pages={350--365},
  year={2009},
  publisher={Elsevier}
}

@article{osella2014concerted,
  title={Concerted control of Escherichia coli cell division},
  author={Osella, Matteo and Nugent, Eileen and Cosentino Lagomarsino, Marco},
  journal={Proceedings of the National Academy of Sciences},
  volume={111},
  number={9},
  pages={3431--3435},
  year={2014},
  publisher={National Academy of Sciences},
  doi = {10.1073/pnas.1313715111}
}

@article{devany2023epithelial,
  title={Epithelial tissue confinement inhibits cell growth and leads to volume-reducing divisions},
  author={Devany, John and Falk, Martin J and Holt, Liam J and Murugan, Arvind and Gardel, Margaret L},
  journal={Developmental cell},
  volume={58},
  number={16},
  pages={1462--1476},
  year={2023},
  publisher={Elsevier}
}

@article{sepulveda2013collective,
  author = {Sep{\'u}lveda, N. and Petitjean, L. and Cochet, O. and Grasland-Mongrain, E. and Silberzan, P. and Hakim, V.},
  title = {Collective Cell Motion in an Epithelial Sheet Can Be Quantitatively Described by a Stochastic Interacting Particle Model},
  journal = {PLoS Computational Biology},
  year = {2013},
  volume = {9},
  number = {3},
  eid = {e1002944},
  doi = {10.1371/journal.pcbi.1002944}
}

@article{chepizhko2018jamming,
  title={From jamming to collective cell migration through a boundary induced transition},
  author={Chepizhko, Oleksandr and Lionetti, Maria Chiara and Malinverno, Chiara and Giampietro, Costanza and Scita, Giorgio and Zapperi, Stefano and La Porta, Caterina AM},
  journal={Soft matter},
  volume={14},
  number={19},
  pages={3774--3782},
  year={2018},
  publisher={Royal Society of Chemistry}
}

@article{armengol2024hydrodynamics,
	title={Hydrodynamics and multiscale order in confluent epithelia},
	author={Armengol-Collado, Josep-Maria and Carenza, Livio Nicola and Giomi, Luca},
	journal={Elife},
	volume={13},
	pages={e86400},
	year={2024}
}

@article{eckert2023hexanematic,
	title={Hexanematic crossover in epithelial monolayers depends on cell adhesion and cell density},
	author={Eckert, Julia and Ladoux, Beno{\^\i}t and M{\`e}ge, Ren{\'e}-Marc and Giomi, Luca and Schmidt, Thomas},
	journal={Nature Communications},
	volume={14},
	number={1},
	pages={5762},
	year={2023},
	publisher={Nature Publishing Group UK London}
}

@article{armengol2023epithelia,
	title={Epithelia are multiscale active liquid crystals},
	author={Armengol-Collado, Josep-Maria and Carenza, Livio Nicola and Eckert, Julia and Krommydas, Dimitrios and Giomi, Luca},
	journal={Nature Physics},
	volume={19},
	number={12},
	pages={1773--1779},
	year={2023},
	publisher={Nature Publishing Group UK London}
}

@article{li2014coherent,
	title={Coherent motions in confluent cell monolayer sheets},
	author={Li, Bo and Sun, Sean X},
	journal={Biophysical journal},
	volume={107},
	number={7},
	pages={1532--1541},
	year={2014},
	publisher={Elsevier}
}

@article{basan2013alignment,
	title={Alignment of cellular motility forces with tissue flow as a mechanism for efficient wound healing},
	author={Basan, Markus and Elgeti, Jens and Hannezo, Edouard and Rappel, Wouter-Jan and Levine, Herbert},
	journal={Proceedings of the National Academy of Sciences},
	volume={110},
	number={7},
	pages={2452--2459},
	year={2013},
	publisher={National Academy of Sciences}
}

@article{barton2017active,
	title={Active vertex model for cell-resolution description of epithelial tissue mechanics},
	author={Barton, Daniel L and Henkes, Silke and Weijer, Cornelis J and Sknepnek, Rastko},
	journal={PLoS computational biology},
	volume={13},
	number={6},
	pages={e1005569},
	year={2017},
	publisher={Public Library of Science San Francisco, CA USA}
}

@article{bi2015density,
	title={A density-independent rigidity transition in biological tissues},
	author={Bi, Dapeng and Lopez, JH and Schwarz, Jennifer M and Manning, M Lisa},
	journal={Nature Physics},
	volume={11},
	number={12},
	pages={1074--1079},
	year={2015},
	publisher={Nature Publishing Group UK London}
}

@article{fletcher2014vertex,
	title={Vertex models of epithelial morphogenesis},
	author={Fletcher, Alexander G and Osterfield, Miriam and Baker, Ruth E and Shvartsman, Stanislav Y},
	journal={Biophysical journal},
	volume={106},
	number={11},
	pages={2291--2304},
	year={2014},
	publisher={Elsevier}
}

@article{richardson2016leader,
	title={Leader cells define directionality of trunk, but not cranial, neural crest cell migration},
	author={Richardson, Jo and Gauert, Anton and Montecinos, Luis Briones and Fanlo, Luc{\'\i}a and Alhashem, Zainalabdeen Mohmammed and Assar, Rodrigo and Marti, Elisa and Kabla, Alexandre and H{\"a}rtel, Steffen and Linker, Claudia},
	journal={Cell reports},
	volume={15},
	number={9},
	pages={2076--2088},
	year={2016},
	publisher={Elsevier}
}

@article{mayor2016front,
	title={The front and rear of collective cell migration},
	author={Mayor, Roberto and Etienne-Manneville, Sandrine},
	journal={Nature reviews Molecular cell biology},
	volume={17},
	number={2},
	pages={97--109},
	year={2016},
	publisher={Nature Publishing Group UK London}
}

@article{trepat2009physical,
	title={Physical forces during collective cell migration},
	author={Trepat, Xavier and Wasserman, Michael R and Angelini, Thomas E and Millet, Emil and Weitz, David A and Butler, James P and Fredberg, Jeffrey J},
	journal={Nature physics},
	volume={5},
	number={6},
	pages={426--430},
	year={2009},
	publisher={Nature Publishing Group UK London}
}

@article{yang2017correlating,
	title={Correlating cell shape and cellular stress in motile confluent tissues},
	author={Yang, Xingbo and Bi, Dapeng and Czajkowski, Michael and Merkel, Matthias and Manning, M Lisa and Marchetti, M Cristina},
	journal={Proceedings of the National Academy of Sciences},
	volume={114},
	number={48},
	pages={12663--12668},
	year={2017},
	publisher={National Academy of Sciences}
}

@article{gov2011moving,
  title={Moving under peer pressure},
  author={Gov, Nir},
  journal={Nature materials},
  volume={10},
  number={6},
  pages={412--414},
  year={2011},
  publisher={Nature Publishing Group UK London}
}

@article{tambe2011collective,
	title={Collective cell guidance by cooperative intercellular forces},
	author={Tambe, Dhananjay T and Corey Hardin, C and Angelini, Thomas E and Rajendran, Kavitha and Park, Chan Young and Serra-Picamal, Xavier and Zhou, Enhua H and Zaman, Muhammad H and Butler, James P and Weitz, David A and others},
	journal={Nature materials},
	volume={10},
	number={6},
	pages={469--475},
	year={2011},
	publisher={Nature Publishing Group UK London}
}

@article{malinverno2017endocytic,
	title={Endocytic reawakening of motility in jammed epithelia},
	author={Malinverno, Chiara and Corallino, Salvatore and Giavazzi, Fabio and Bergert, Martin and Li, Qingsen and Leoni, Marco and Disanza, Andrea and Frittoli, Emanuela and Oldani, Amanda and Martini, Emanuele and others},
	journal={Nature materials},
	volume={16},
	number={5},
	pages={587--596},
	year={2017},
	publisher={Nature Publishing Group UK London}
}

@article{park2015unjamming,
  title={Unjamming and cell shape in the asthmatic airway epithelium},
  author={Park, Jin-Ah and Kim, Jae Hun and Bi, Dapeng and Mitchel, Jennifer A and Qazvini, Nader Taheri and Tantisira, Kelan and Park, Chan Young and McGill, Maureen and Kim, Sae-Hoon and Gweon, Bomi and others},
  journal={Nature materials},
  volume={14},
  number={10},
  pages={1040--1048},
  year={2015},
  publisher={Nature Publishing Group UK London}
}

@article{petitjean2010velocity,
  title={Velocity fields in a collectively migrating epithelium},
  author={Petitjean, Laurence and Reffay, Myriam and Grasland-Mongrain, Erwan and Poujade, Mathieu and Ladoux, Beno{\i}t and Buguin, Axel and Silberzan, Pascal},
  journal={Biophysical journal},
  volume={98},
  number={9},
  pages={1790--1800},
  year={2010},
  publisher={Elsevier}
}

@article{reffay2011orientation,
  title={Orientation and polarity in collectively migrating cell structures: statics and dynamics},
  author={Reffay, Myriam and Petitjean, Laurence and Coscoy, Sylvie and Grasland-Mongrain, Erwan and Amblard, Francois and Buguin, Axel and Silberzan, Pascal},
  journal={Biophysical journal},
  volume={100},
  number={11},
  pages={2566--2575},
  year={2011},
  publisher={Elsevier}
}

@article{ladoux2016front,
  title={Front--rear polarization by mechanical cues: from single cells to tissues},
  author={Ladoux, Benoit and M{\`e}ge, Ren{\'e}-Marc and Trepat, Xavier},
  journal={Trends in cell biology},
  volume={26},
  number={6},
  pages={420--433},
  year={2016},
  publisher={Elsevier}
}

@article{devany2021cell,
  title={Cell cycle--dependent active stress drives epithelia remodeling},
  author={Devany, John and Sussman, Daniel M and Yamamoto, Takaki and Manning, M Lisa and Gardel, Margaret L},
  journal={Proceedings of the National Academy of Sciences},
  volume={118},
  number={10},
  pages={e1917853118},
  year={2021},
  publisher={National Academy of Sciences},
doi = {10.1073/pnas.1917853118}
}

@article{trepat2011plithotaxis,
  title={Plithotaxis and emergent dynamics in collective cellular migration},
  author={Trepat, Xavier and Fredberg, Jeffrey J},
  journal={Trends in cell biology},
  volume={21},
  number={11},
  pages={638--646},
  year={2011},
  publisher={Elsevier},
  doi = {10.1016/j.tcb.2011.06.006.}
}

@article{trepat2018mesoscale,
  title={Mesoscale physical principles of collective cell organization},
  author={Trepat, Xavier and Sahai, Erik},
  journal={Nature Physics},
  volume={14},
  number={7},
  pages={671--682},
  year={2018},
  publisher={Nature Publishing Group UK London}
}

@article{li2025emergence,
  title={Emergence of cellular nematic order is a conserved feature of gastrulation in animal embryos},
  author={Li, Xin and Huebner, Robert J and Williams, Margot LK and Sawyer, Jessica and Peifer, Mark and Wallingford, John B and Thirumalai, D},
  journal={Nature Communications},
  volume={16},
  number={1},
  pages={5946},
  year={2025},
  publisher={Nature Publishing Group UK London}
}

@article{comba2022spatiotemporal,
  title={Spatiotemporal analysis of glioma heterogeneity reveals COL1A1 as an actionable target to disrupt tumor progression},
  author={Comba, Andrea and Faisal, Syed M and Dunn, Patrick J and Argento, Anna E and Hollon, Todd C and Al-Holou, Wajd N and Varela, Maria Luisa and Zamler, Daniel B and Quass, Gunnar L and Apostolides, Pierre F and others},
  journal={Nature communications},
  volume={13},
  number={1},
  pages={3606},
  year={2022},
  publisher={Nature Publishing Group UK London}
}

@article{jimenez2016curvature,
  title={Curvature-controlled defect localization in elastic surface crystals},
  author={Jim{\'e}nez, Francisco L{\'o}pez and Stoop, Norbert and Lagrange, Romain and Dunkel, J{\"o}rn and Reis, Pedro M},
  journal={Physical review letters},
  volume={116},
  number={10},
  pages={104301},
  year={2016},
  publisher={APS}
}

\end{document}